# Stabilization and current-induced motion of antiskyrmion in the presence of anisotropic Dzyaloshinskii-Moriya interaction

Siying Huang[1], Chao Zhou[1], Gong Chen[2]*, Hongyi Shen[1], Andreas K. Schmid[3], Kai Liu[2], and Yizheng Wu[1,4]*

[1]Department of Physics and State Key Laboratory of Surface Physics, Fudan University, Shanghai 200433, People's Republic of China

[2]Physics Department, University of California, Davis, California 95616, USA

[3]NCEM, Molecular Foundry, Lawrence Berkeley National Laboratory, Berkeley, California 94720, USA

[4]Collaborative Innovation Center of Advanced Microstructures, Nanjing 210093, China

Email：gchenncem@gmail.com, wuyizheng@fudan.edu.cn

Abstract：
Topological defects in magnetism have attracted great attention due to fundamental research interests and potential novel spintronics applications. Rich examples of topological defects can be found in nanoscale non-uniform spin textures, such as monopoles, domain walls, vortices, and skyrmions. Recently, skyrmions stabilized by the Dzyaloshinskii-Moriya interaction have been studied extensively. However, the stabilization of antiskyrmions is less straightforward. Here, using numerical simulations we demonstrate that antiskyrmions can be a stable spin configuration in the presence of anisotropic Dzyaloshinskii-Moriya interaction. We find current-driven antiskyrmion motion that has a transverse component, namely antiskyrmion Hall effect. The antiskyrmion gyroconstant is opposite to that for skyrmion, which allows the current-driven propagation of coupled skyrmion-antiskyrmion pairs without apparent skyrmion Hall effect. The antiskyrmion Hall angle strongly depends on the current direction, and a zero antiskyrmion Hall angle can be achieved at a critic current direction. These results open up possibilities to tailor the spin topology in nanoscale magnetism, which may be useful in the emerging field of skyrmionics.

# I. Introduction

Topological defects such as monopoles, domain walls, vortices, and skyrmions play crucial roles in low dimensional magnetic systems [1], where topological features dominate their magnetic and dynamic properties. Examples include the switching of vortex core polarity via creation/annihilation of vortex - antivortex pair [2], magnetization reversal via soliton pair creation [3], topological protection of skyrmions [4] and domain walls [5]. Among these spin textures, magnetic skyrmions [6,7,8] have attracted increasing attention owing to the ultralow current density threshold to move them [9,10,11,12].

Magnetic skyrmions are nanometer-sized circular quasiparticles that have homo-chiral spin textures [4,12], where the nature of the magnetic chirality can be realized by several mechanisms, such as the antisymmetric exchange interaction, namely the Dzyaloshinskii-Moriya interaction (DMI) [13,14,15,16,17], four spin interactions [18], artificial confinement in patterned structures [19,20,21], or the competing exchange interactions [22.23]. In case of the DMI, broken inversion symmetry of the system is required as the interaction vanishes in symmetric systems. The DMI energy term can be written as $E_{\mathrm{DM}} = -\mathbf{D}_{ij} \cdot (\mathbf{S}_i \times \mathbf{S}_j)$, where $\mathbf{S}_i$ and $\mathbf{S}_j$ are two spins on neighbouring atomic sites $i$ and $j$, and $\mathbf{D}_{ij}$ is the vector characterizing the DMI. Magnetic skyrmions have been observed experimentally in non-centrosymmetric bulk magnets [4,9,10,24,25], ultra-thin films where the inversion symmetry breaks at the interface [15,26,27,28,29,30], and artificial structures [19,20,21].

Experimental observations of magnetic skyrmion states have motivated studies on their creation [27,31], manipulation [31,32,33] and electric detection [34] in the presence of electric current [9,10,11,27,31,32] as well as magnetic field [9,10,11,27,28,29,30,31]. For possible skyrmionic memory and logic device applications envisioned so far, current-driven propagation of magnetic skyrmions is highly relevant to device performance, with two key parameters being the critical current for skyrmion propagation and the maximum propagation speed. Theoretical studies predict that skyrmions propagate along trajectories away from the current direction due to the Magnus force [4,35,36,37,38,39,40,41,42,43,44], following the so-called skyrmion Hall effect, which has been recently observed in Néel-type skyrmions in thin film systems [45,46]. Theoretical studies have further proposed that the skyrmion Hall effect can be suppressed by zeroing the net topological charge in two coupled skyrmions with opposite topological charges, either in antiferromagnetically-coupled trilayers with skyrmions on both sides [47], or in skyrmions in antiferromagnetic materials [48].

Magnetic antiskyrmions are topologically non-trivial chiral spin quasiparticles that may occur in cases where the magnetic chirality is anisotropic (as opposed to skyrmions with isotropic in-plane chirality). They have received less attention up to now [4,49,50,51,52]. These theoretical works predict that antiskyrmions can be stabilized in materials belonging to the crystallographic class $D_{2d}$ [53], as well as in frustrated exchange interaction systems [50] and in dipolar magnets [52].

In this work, we focus on the energetics of magnetic antiskyrmions in the presence of anisotropic interfacial DMI, and further explore their electric current-driven dynamics. We provide a phase diagram of spin configurations including skyrmions, antiskyrmions and multidomains (spin spirals) in a 2D space of the DMI vector along two orthogonal in-plane directions. We show that the topological charge of the antiskyrmions may also induce a Magnus force-associated propagation deviation, i.e. an antiskyrmion Hall effect. Moreover, we find that the anisotropic spin texture of antiskyrmions gives rise to significant anisotropy in the skyrmions' response to applied electric current, i.e. the antiskyrmion Hall angle (between propagation direction and current direction) strongly depends on the applied current direction with respect to the internal spin texture of antiskyrmions. Indeed the antiskyrmion Hall angle ranged from positive to negative, even crossing zero degree for certain current directions. This tunability of the antiskyrmion Hall angle is a new degree of freedom that enables the control of trajectories of topological quasiparticles, which might be useful in skyrmionics-based memory or logic devices. We also investigate the current-driven dynamics of a coupled skyrmion-antiskyrmion pair in which the Magnus force of the skyrmion and antiskyrmion can be cancelled.

## II. Stabilization of antiskyrmions by anisotropic DMI

We first discuss differences in the mechanisms that stabilize skyrmions and antiskyrmions. At the interface between a thin magnetic film and a heavy metal adjacent layer the DM vector $\mathbf{D}_{ij}$ between spins $\mathbf{S}_i$ and $\mathbf{S}_j$ on atomic sites $i$ and $j$ usually lies within the film plane and normal to the distance vector $\mathbf{r}_{ij}$ [17,18]. Note that the orientation of the $\mathbf{D}_{ij}$ vector plays a critical role in determining the chiral spin configuration. Generally, skyrmion configurations result from isotropic interfacial DMI [17,18,26,27,28,29,30,31], i.e. in systems where the $\mathbf{D}_{ij}$ vector is either parallel or antiparallel to $\mathbf{z} \times \mathbf{r}_{ij}$, where $\mathbf{z}$ is interface normal direction [17]. For instance, in a four-fold symmetric system such as an fcc(001) interface, four $\mathbf{D}_{ij}$ vectors at four $j$ sites ($j = 1,2,3,4$) adjacent to an atom at site $i$ (see Fig. 1(a)) can be written as $\mathbf{D}_1 = D_1\hat{\mathbf{y}}$, $\mathbf{D}_2 = -D_2\hat{\mathbf{x}}$, $\mathbf{D}_3 = -D_3\hat{\mathbf{y}}$, $\mathbf{D}_4 = D_4\hat{\mathbf{x}}$, respectively, where $D_j$ is strength of the DMI vector on $j$ sites, and $\hat{\mathbf{x}}$ and $\hat{\mathbf{y}}$ correspond to the unit vector along the *x* and *y* axis, respectively. The DMI vectors on opposite $j$ sites are opposite, i.e. $\mathbf{D}_1 = -\mathbf{D}_3$, and $\mathbf{D}_2 = -\mathbf{D}_4$, so for simplicity we discuss the DMI vector configurations in $D_1 - D_2$ space in the rest of the paper. Most previous studies usually assumed that the interfacial DMI vectors on four $j$ sites have the same rotational sense [17], e.g. the DMI vector configuration shown in Fig. 1(a), resulting in same chirality along all in-plane directions (Fig. 1(c)). Note that flipping the sign of all DMI vectors will reverse the magnetic chirality within skyrmions, but the topological charge remains the same for both chiralities when the cores of skyrmions point in the same direction (Fig. 1(e)) [4].

In most previous studies it has always been assumed that the strengths of $D_1$ and $D_2$ are equal, a view that is generally accepted in systems with four-fold symmetry and in polycrystalline systems. In contrast, atomic configurations with broken in-plane rotation symmetry at the interface may lead to anisotropic interfacial DMI [17]. For instance, first principles calculations predict that the bcc Fe/W(110) interface has opposite $\mathbf{D}_{ij}$ vectors along two orthogonal in-plane directions [001] and [1$\bar{1}$0] [54]. Recently, anisotropic DMI was experimentally observed at the Co/W(110) interface, where the strength of the DMI $\mathbf{D}_{ij}$ vector was found to have the same sign but 2.5 times stronger magnitude along the bcc [1$\bar{1}$0] direction compared to the orthogonal direction bcc [001] [55]. Note that the in-plane anisotropy of the DMI vector $\mathbf{D}_{ij}$ changes the energy landscape of chiral spin textures along different in-plane direction. For instance antiskyrmions are stabilized when the $\mathbf{D}_{ij}$ vectors along *x* and *y* directions have opposite sign, i.e. along the vertical direction in the sketch in Fig. 1(b) the vectors remain the same as in Fig. 1(a) ($\mathbf{D}_1 = D_1\hat{\mathbf{y}}$), but along the horizontal direction vectors flip their sign ($\mathbf{D}_2 = D_2\hat{\mathbf{x}}$ instead of $\mathbf{D}_2 = -D_2\hat{\mathbf{x}}$). This anisotropic DMI configuration favours opposite chirality along *x* and *y* direction, which allows the stabilization of magnetic antiskyrmions (Fig. 1(d)). Note that the energies of dipolar interaction between skyrmions and antiskyrmions are slightly different due to the different length fraction of Néel- vs. Bloch-type domain walls on their boundaries. However the dipolar energies of opposite Néel-type (or Bloch-type) chiralities are degenerate [56,57], therefore dipolar interactions do not influence the anisotropic chirality of the antiskyrmions. Note also that the antiskyrmion shown in Fig. 1(f) carries the opposite topological charge compared to the skyrmion sketched in Fig. 1(e) [4].

Next we discuss the possibility to stabilize antiskyrmions in nanodisks in the presence of anisotropic interface DMI. We performed micromagnetic simulations using the OOMMF code including the DMI [58,59], and numerically calculated the relaxed micromagnetic state of 80-nm-wide 0.4nm thick nanodisks in zero field. As described in more detail in the Methods section, our model captures the DMI strengths of $D_1$ and $D_2$ in terms of the area unit energy densities $D_x$ and $D_y$. In the simulations, plausible initial states of antiskyrmions, quasi-uniform and multidomain magnetization distributions are relaxed to obtain different final states (see Methods). We first assume that the DMI vectors $(D_x, D_y)$ have the same magnitude and opposite sign, i.e. $D_x = -D_y$. Figure 2(a) shows the total micromagnetic energy of three possible configurations as a function of $D_x$, including a quasi-uniform state (red circles), antiskyrmion state (black squares) and 2π-rotation state (blue triangles, explained below) [60]. It is clear that for $D_x < D'$, the most stable state is quasi-uniform state with titled spins at the edge, consistent with results reported in Ref. 60. The antiskyrmion state becomes the ground state when $D' < D_x < D"$, where the size of antiskyrmions increases with the magnitude of $D_x$. For $D_x > D"$, the most favoured state is a 2π-rotation state with the spins rotating by 2π from the center to the edge of the disk, and for stronger $D_x$, nπ-rotation states with $n > 2$ may exist. Skyrmion textures with isotropic chirality are not discussed here because their DMI energy would far exceed that of antiskyrmions in the anisotropic DMI case. Figure 2(b) shows a typical spin configuration of a stable antiskyrmion with $D_x = -D_y = 4 \text{ mJ m}^{-2}$.

### III. Simulated spin texture phase diagram in an anisotropic DMI system

It is interesting to explore further the spin configurations of such nanodisks under conditions where $D_x$ and $D_y$ are unbalanced [54,55]. A phase diagram shown in Fig. 3(a) summarizes the dependence of spin texture ground states of the disk in $(D_x, D_y)$ space (see details in Methods), where the $(D_x, D_y)$ quadrant represents isotropic chirality states containing skyrmion configurations, and the $(D_x, -D_y)$ quadrant represents anisotropic chirality states containing antiskyrmion configurations. The shapes of skyrmions (cyan region in the phase diagram) and antiskyrmions (yellow region) evolve from circular (Fig. 3(b) and 3(c)) to elliptical (Fig. 3(d) and 3(e)) when the magnitude of $D_x$ and $D_y$ are unequal. Note that such elliptical deformation may also be present in a biaxial magnetic anisotropy system [61]. The quasi-uniform out-of-plane ferromagnetic states (darker cyan and darker yellow regions) have tilted spins at the edge due to the presence of the DMI [60]. Depending on the magnitude of $D_x$ and $D_y$, multi-domain ground states either form stripe-like phases (Fig. 3(f) and 3(g)) or $n\pi$-rotation phases (Fig. 3(h) and 3(i) shows the spin texture with $n = 2$),

## IV. Current-induced motion of antiskyrmions in nanotracks

Exploring the dynamical properties of antiskyrmions is fundamentally interesting, as these objects are a unique type of topologically charged quasiparticles. In this section we explore, by micromagnetic simulations, current-induced motions of skyrmions and antiskyrmions confined within nanotracks. For these simulations we implemented in the OOMMF code additional torque terms added to the LLG equation (see Methods) [58,59]. A spin current polarized along $+y$ direction is injected vertically by the spin Hall effect [11,60]. We simulated both skyrmions, under isotropic DMI $(D_x, D_y)$, and antiskyrmions with equivalent size under anisotropic DMI $(D_x, -D_y)$. To highlight the importance of topology, the magnitudes of $D_x$ and $D_y$ are always kept to $3\ mJ\ m^{-2}$ in this section. Simulations show that skyrmions and antiskyrmions propagate with the same velocity under a driving current density $j = 1 \text{MA cm}^{-2}$ (Fig. 4(a) and Supplementary Movie 1 in ref. 62), and their velocities parallel to the nanotrack increase linearly with the current density. With higher current density, both skyrmion and antiskyrmion show a transverse velocity in the y-direction.

The transverse motions for skyrmion and antiskyrmion are in opposite directions. This is understood within the picture of the Thiele equation as used to describe skyrmion Hall effect in rigid skyrmion systems [4,38,43,44,43,45],

$$\mathbf{G} \times \mathbf{v} - \alpha \mathbf{D} \cdot \mathbf{v} - 4\pi \mathbf{B} \cdot \mathbf{j} = 0 \tag{1}$$

where $G = (0,0,-4\pi Q)$ is the gyromagnetic coupling vector, and the skyrmion and antiskyrmion have opposite topological charge $Q$; $\mathbf{v} = (v_x, v_y)$ is the propagation velocity along the $x$ and $y$ axis, respectively; $\alpha$ is the Gilbert damping coefficient, and $\mathbf{D}$ is the dissipative force tensor. The tensor $\mathbf{B}$ represents the efficiency of the spin Hall torque over the skyrmion/antiskyrmion, and $\mathbf{j}$ is the electric current density flowing in the heavy metal. The first term in Eq. (1) is the topological Magnus force [41,45], which induces the transverse motion of skyrmions (or anitskyrmion) with respect to the driving current. The second term is the dissipative force due to the magnetic damping of a moving magnetic skyrmion (or antiskyrmion), and the third term shows the driving force from the spin Hall torque. The Thiele equation yields $v_x = \frac{-j\alpha \text{DB}_{xx}}{(\alpha \text{D})^2 + Q^2}$ and $v_y = \frac{jQ\text{B}_{xx}}{(\alpha \text{D})^2 + Q^2}$ for the velocity components of the skyrmion or antiskyrmion along $x$ and $y$ direction (see Methods), showing that the opposite transverse motion $v_y$ of skyrmion and antiskyrmion is due to their opposite topological charge $Q$. Such transverse motion stops near the edge of the nanotrack due to the repulsive interaction caused by the tilting magnetization at the edge induced by DMI [11,35]. (The propagation without the edge effect will be discussed in next section.) Just as the topological properties of skyrmions can reduce the influence of defects on their motion, our simulations suggest that antiskyrmions are equally protected from defect-induced perturbations by their topological order (see Supplementary Figure 1 and movie 2 in ref. 62). When the driving current is sufficiently strong, the transverse force on either skyrmions or antiskyrmions is sufficient to overcome the repulsive barrier at the track edge, resulting in annihilation. In the case of antiskyrmions this annihilation occurs earlier than in the case of skyrmion under the similar conditions (Fig. 4(a)). This is because antiskrymions can be slightly rotated by the spin-orbit torque during their motion – consequently, when antiskyrmions approach the track-edge as a result of strong spin current, the radial component of magnetization near the track edge is reduced and causes a weaker repulsive interaction between the antiskyrmion and the edge.

## V. Anisotropic antiskyrmion Hall effect.

Inspired by theoretical predictions and experimental observations of the skyrmion Hall effect [4,35,36,37,38,39,40,41,42,45,46,47,48], we investigate whether antiskyrmions may also exhibit current-driven transverse motion associated with their anisotropic spin texture. Figures 5(a) and 5(b) show the typical propagation trajectories of a single skyrmion and antiskyrmion stabilized by identical DMI vector magnitudes (the amplitude of both $D_x$ and $D_y$ are set to $3\text{ mJ m}^{-2}$) in the presence of current density of $j = 10\text{ MA cm}^{-2}$ along the $x$ direction (see Supplementary Movies 3 and 4 in ref. 62). One feature is that, without the edge effect, the transverse motions of skyrmions and antiskyrmions are in opposite directions (along $y$ axis) while they are both propagating along $+x$ direction. These transverse motions can be quantified as the skyrmion/antiskyrmion Hall angle, i.e. the relative angle from the applied current to the motion trajectory direction, which can be calculated from the ratio of $v_y/v_x$. We found that both the skyrmion and antiskyrmion Hall angles are equal to $arctan(-\frac{Q}{\alpha D})$, thus they have the opposite values due to the opposite topological charge $Q$, and the value increases with the DMI strength, supported by both calculation and simulation (Fig. 5(e)).

In contrast to skyrmions in which the spin texture is isotropic, antiskyrmions have anisotropic in-plane spin textures. In the following we explore the possible anisotropic responses when current is injected in arbitrary directions $\theta$ with respect to the $+x$ axis. The $\theta$ angle dependent skyrmion Hall angle and antiskyrmion Hall angle can also be understood in the picture of the modified Thiele equation. The derived skyrmion Hall angle equals to $arctan(-\frac{Q}{\alpha D})$, independent of $\theta$, whereas the antiskyrmion Hall angle equals $arctan\left(-\frac{Q}{\alpha D}\right) - 2\theta$ (Methods), so the antiskyrmion Hall angle rotates oppositely against the rotation of angle $\theta$. Simulation in Fig. 5(c) indeed shows that the antiskyrmion Hall angle changes its sign (being positive with respect to current direction) when the current is injected along the $-y$ direction ($\theta = -90°$) (see Supplementary Movie 5 in ref. 62). This $\theta$-dependent sign change of the antiskyrmion Hall angle gives rise to opportunities to tailor the trajectories of antiskyrmions in the presence of electric current. Moreover, the transverse motion of antiskyrmions can be eliminated when current is injected at the angle $\theta = arctan\left(-\frac{Q}{\alpha D}\right)/2$. This allows for the propagation of antiskyrmions along the current direction with zero Hall effect, which is supported by the simulation shown in Fig. 5(d) and Supplementary Movie 6 in ref. 62, where the $\theta$ angle equals to $-34.5°$. Figure 5(f) shows the simulated linear dependence of antiskyrmion Hall angle on $\theta$ with current density of $j = 10 \text{ MA cm}^{-2}$ (see additional angle dependence in Supplementary Moives 7 and 8 in ref. 62), which can be well fitted by the antiskyrmion Hall angle of $arctan\left(-\frac{Q}{\alpha D}\right) - 2\theta$ derived from the Thiele equation. The total velocity of antiskyrmion $v = \frac{B_{xx}}{\sqrt{(\alpha D)^2 + Q^2}}|\mathbf{j}|$ is independent of the angle $\theta$. We found that the critical current direction with zero antiskyrmion Hall angle depends on the DMI strength, and also shows little current dependence for $j < 20$ MA cm$^{-2}$, but could change obviously at larger current density and reach $-33.1°$ with $j = 50$ MA cm$^{-2}$. This is because the spin configuration in anitskyrmions can be modified by the spin torque under strong spin current, resulting in the size change of antiskyrmions (see Supplementary Figure 2 in ref. 62).

We expect that this unique property could significantly benefit the design of spintronic devices based on skyrmions and antiskyrmions. As shown in Fig. 4(b), the maximum velocity of a single skyrmion or antiskyrmion is typically less than 100 m s$^{-1}$, limited by the competition between the skyrmion/antiskyrmion Hall effect and the edge confining force [11,37,47]. By patterning a nanotrack along the direction enabling zero antiskyrmion Hall angle as shown in Fig. 5(d), the maximum velocity of antiskyrmions can potentially be greatly increased, allowing a faster and denser design of data technologies based on anitskyrmions.

**VI. Coupled antiskyrmion-skyrmion pairs without skyrmion Hall effect.**

Due to the skyrmion Hall effect, magnetic skyrmions may be annihilated at the edges of nanotracks in the presence of a significant current (Fig. 4(a)), this effect limits the skyrmion motion speed in nanotracks. It was proposed that the skyrmion Hall effect can be efficiently suppressed by building pairs of antiferromagnetically exchange-coupled skyrmions in trilayer systems, where the Magnus forces on the two skyrmions are cancelled due to opposite topological charge $Q$ [47]. Considering the opposite $Q$ for skyrmion and antskyrmion, we propose that the transverse motion of a current-driven ferromagnetically coupled antiskyrmion-skyrmion pair can be eliminated as well (see the sketch of the structure in Fig. 6(a)). When coupled by interface ferromagnetic exchange interaction across the spacer layer, both the antiskyrmion in the top layer and the skyrmion in the bottom layer can be stabilized in the track (Methods). Note that the calculated sizes of the coupled antiskyrmion and skyrmion are equal, but decrease with the interface exchange energy $\sigma$. When $\sigma$ increases up to $0.12$ mJ m$^{-2}$, the coupled antiskyrmion-skyrmion pair becomes unstable (see Supplementary Figure 3 in ref. 62).

We injected a current with the density $j = 5$ MA cm$^{-2}$ in the track. With weak interlayer coupling for $\sigma < 0.015$ mJ m$^{-2}$, antiskyrmion and skyrmion move along the track with the opposite transverse motion due to the opposite topological charge (see Fig. 6(c) and Supplementary Movie 9 in ref. 62). Because of the interface exchange coupling, the angle between their trajectories and $x$ direction has been significantly reduced compared to the skyrmion Hall angle in the single skyrmion configuration. While skyrmion and antiskyrmion move more apart from each other, they eventually decouple and move to track edges separately. Our simulation reveals that the decoupling of the skyrmion-antiskyrmion pair happens later for stronger $\sigma$, and the decomposition distance of the skyrmion-antiskyrmion pair has a dramatic increase for $\sigma \sim 0.015$ mJ m$^{-2}$ with a current density of $j = 5$MA cm$^{-2}$, as shown in Fig. 6(b). For $\sigma > 0.016$ mJ m$^{-2}$, the antiskyrmion-skyrmion pair will move parallel along the applied current direction synchronously (Fig. 6(d) and Supplementary Movie 10 in ref. 62). As shown in Fig. 4b, the maximum velocity of a single skyrmion or antiskyrmion is typically less than $10^2$ m s$^{-1}$, limited by the edge confining force. The maximum velocity of a coupled skyrmion-antiskyrmion pair is much larger than a single skyrmion or antiskyrmion, and the pair can move along the central line of the nanotrack at a high speed of a few hundred meters per second, making this system a good candidate to utilize the skyrmions or antiskyrmions in confined geometries.

**VII. Conclusion**

In summary, we have studied the effect of anisotropic antisymmetric exchange interaction on the chiral spin textures by numerical simulations. We find that magnetic antiskyrmions can be stabilized in nanodisks in the presence of anisotropic interfacial DMI. A phase diagram of chiral spin textures is shown, including skyrmion and antiskyrmion in the disk, in the absence of applied magnetic field. Current-driven propagation of antiskyrmions contains a transverse component due to the Magnus force associated with the topological charge, opposite to the transverse motion of skyrmions. Moreover, the antiskyrmion Hall angle strongly depends on the direction of the applied current with respect to the in-plane spin structures of antiskyrmions. This constitutes a new degree of freedom to manipulate the trajectories of chiral spin textures, and enables the design where antiskyrmion propagation is free of transverse motion. We further show that coupling an antiskyrmion-skyrmion pair in a trilayer structure is an alternative approach to suppress the topological charge-associated Magnus forces. These results may trigger experimental efforts to explore antiskyrmions, and the current-driven dynamics of antiskyrmions has exciting potentials for novel functionalities.

## VIII. Methods

### A. Micromagnetic modelling

In this study, we considered a 0.4-nm-thick perpendicularly magnetized cobalt film grown on a substrate inducing anisotropic DMI. In a continuous magnetization model, the DMI energy can be expressed as:

$$\varepsilon_{DM} = D_x \left( m_z \frac{\partial m_x}{\partial x} - m_x \frac{\partial m_z}{\partial x} \right) + D_y \left( m_z \frac{\partial m_y}{\partial y} - m_y \frac{\partial m_z}{\partial y} \right) \quad (2)$$

where $m_x, m_y, m_z$ are the components of the unit magnetization. $D_x$ and $D_y$ are area unit energy densities related to the strength of $D_1$ and $D_2$ shown in Fig. 1, with a $1/at$ proportional factor, where $a$ is the atomic lattice constant and $t$ is film thickness [11].

## B. Simulations of the ground state

The simulations of this finite micromagnetic system were done using the modified OOMMF code including the anisotropic DMI [58,59]. We first investigated the relaxed state of an 80nm wide, 0.4nm thick nanodisk on a substrate in zero field. We used perpendicular magnetic anisotropy $K = 0.6$ MJ $m^{-2}$ along the $z$ axis, exchange stiffness $A = 1.5$ pJ m$^{-1}$, Gilbert damping $\alpha = 0.3$ and saturation magnetization $M_s = 580$ kA m$^{-1}$, similar to the parameters used in Ref. 11. The unit cell in the simulation is $0.4 \times 0.4 \times 0.4$ nm$^3$. To obtain the total energies of the quasi-uniform states, antiskyrmion states and $2\pi$-rotation states as a function of $D_x$, as summarized in Fig. 2(a), we first set initial spin configurations corresponding to these states and then relax the system to the final states. To obtain the phase diagram in $(D_x, D_y)$ space shown in Fig. 3(a), we initially set the magnetization in the central 20nm diameter region along the $+z$ direction and set the magnetization outside of the 20nm diameter along the $-z$ direction, and then relaxed the magnetic system from the initial state by solving Landau-Lifschitz-Gilbert (LLG) equation step-by-step.

## C. Simulations of current-driven skyrmion/antiskyrmion propagations

To simulate current-driven motions of skyrmions and antiskyrmions by the vertical injection of spin current (as it occurs by spin Hall effect in magnetic films grown on heavy metals with large spin Hall angle), we used the OOMMF code including extra torques terms added to the LLG equation [11,59]:

$$\boldsymbol{\tau}_1 = -\gamma \tau_1 (\hat{\mathbf{m}} \times \hat{\boldsymbol{\sigma}} \times \hat{\mathbf{m}}) \tag{3}$$

$$\boldsymbol{\tau}_2 = -\gamma \tau_2 (\hat{\mathbf{m}} \times \hat{\boldsymbol{\sigma}}) \tag{4}$$

where $\boldsymbol{\tau}_1$ is the in-plane torque and $\boldsymbol{\tau}_2$ is the out-of-plane torque with $\tau_1$ and $\tau_2$ being torque magnitudes. $\hat{\mathbf{m}}$ is the unit vector along the magnetization axis, $\hat{\boldsymbol{\sigma}}$ is direction of the current polarization vector and $\gamma$ is the gyromagnetic ratio. As shown in Ref. 11, for films thinner than the absorption length of spin transfer, incomplete spin transfer of the injected spin current can be taken into account by a renormalized spin polarization P, and we used the value of P=40% in our simulations. In simulations of current-induced motion of skyrmions and antiskyrmions, the unit cell is $1 \times 1 \times 0.4$ nm$^3$.

For current-induced motion in nanotracks, the cross-section of the nanotrack is 40 nm × 0.4 nm. To obtain antiskyrmions and skyrmions with reduced diameters, the anisotropy $K$ in this simulation was set to 0.8 MJ m$^{-2}$. A spin current polarized along the +y direction is injected from the heavy metal layer along the z direction to the Co layer due to the spin Hall effect [11,60]. To simulate the antiskyrmion Hall effect under various applied current directions, we used a sufficiently large square thin film (300 nm × 300 nm × 0.4 nm) to avoid edge effects. The skyrmion (or antiskyrmion) is initially positioned at the center of the track to reduce the influence from the edge magnetization. In the simulation of current-driven antiskyrmion-skyrmion pair motion, we used a trilayer track of 80 nm × 1.2 nm cross section (the thicknesses of top layer, spacer layer and bottom layer are all 0.4 nm). The increased width of 80 nm is used here to reduce the influence of track edges so that the transverse motions can be clearly observed. In the top layer, the anisotropic DMI is set to $D_x = -D_y = 3$ mJ $m^{-2}$, and in the bottom layer the isotropic DMI is set to $D_x = D_y = 3$ mJ $m^{-2}$.

### D. Current-direction dependent antiskyrmion Hall effect

To understand the anisotropic skyrmion/antiskyrmion Hall angle, we start from the modified Thiele equation $\mathbf{G} \times \mathbf{v} - \alpha \mathbf{D} \cdot \mathbf{v} - 4\pi \mathbf{B} \cdot \mathbf{j} = 0$, where the dissipative force tensor $\mathbf{D}$ is determined by the spin configuration in skyrmion or antiskyrmion, which is given by $D = D_{xx} = D_{yy} =$

$\int_{UC}(\partial_i \mathbf{m} \cdot \partial_j \mathbf{m})\,dx\,dy$ and $D_{xy} = D_{yx} = 0$ for both skyrmion and antiskyrmion configurations. The tensor $\mathbf{B} = \begin{pmatrix} B_{xx} & 0 \\ 0 & B_{yy} \end{pmatrix}$ can be determined by the detailed spin configuration and the topological charge $Q$, with $B_{xx} = B_{yy}$ for skyrmion and $B_{xx} = -B_{yy}$ for antiskyrmion. $\boldsymbol{v} = (v_x, v_y)$ is the drift velocity of skyrmion or antiskyrmion along the x and y axis, respectively. $\boldsymbol{j} = (j\cos\theta, j\sin\theta,)$ is the electrical current density flowing in the heavy metal, where $j$ is the magnitude of the applied current density and $\theta$ is the relative angle between the current and the x axis. Now the Thiele equation yields

$$\begin{cases} v_x = \frac{j}{(\alpha D)^2 + Q^2}(-\alpha D B_{xx}\cos\theta - QB_{yy}\sin\theta) \\ v_y = \frac{j}{(\alpha D)^2 + Q^2}(QB_{xx}\cos\theta - \alpha D B_{yy}\sin\theta) \end{cases} \quad (5)$$

Therefore the skyrmion Hall angle and antiskyrmion Hall angle can be calculated by the ratio of $v_y/v_x$, where the skyrmion Hall angle is $arctan(-\frac{Q}{\alpha D})$, and the antiskyrmion Hall angle is $arctan\left(-\frac{Q}{\alpha D}\right) - 2\theta$ (see details in Supplementary Note 1 in ref. 62). The total velocity for both skyrmion and antiskyrmion is $v = \frac{B_{xx}}{\sqrt{(\alpha D)^2 + Q^2}}j$.

*Note added*. After the original submission of this work, two relevant studies have appeared. Nayak *et al.* [63] reported an observation of magnetic antiskyrmions above room temperature in tetragonal Heusler materials with $D_{2d}$ symmetry, and Hoffmann et al. [64] reported the possibility of stabilizing antiskyrmions in (110) oriented ultrathin films with $C_{2v}$ symmetry.

# Acknowledgements

This work was supported by the National Key Basic Research Program of China (Grant No. 2015CB921401), National Key Research and Development Program of China (Grant No.

2016YFA0300703), National Natural Science Foundation of China (Grants No. 11474066 and No. 11434003), and the Program of Shanghai Academic Research Leader (No. 17XD1400400). Work at UCD was supported by the UC Office of the President Multicampus Research Programs and Initiatives (MRP-17-454963) (G.C.) and the US NSF (DMR-1610060) (K.L.). Work at LBL was supported by the Office of Science, Office of Basic Energy Sciences, Scientific User Facilities Division, of the U.S. Department of Energy under Contract No. DE-AC02—05CH11231.

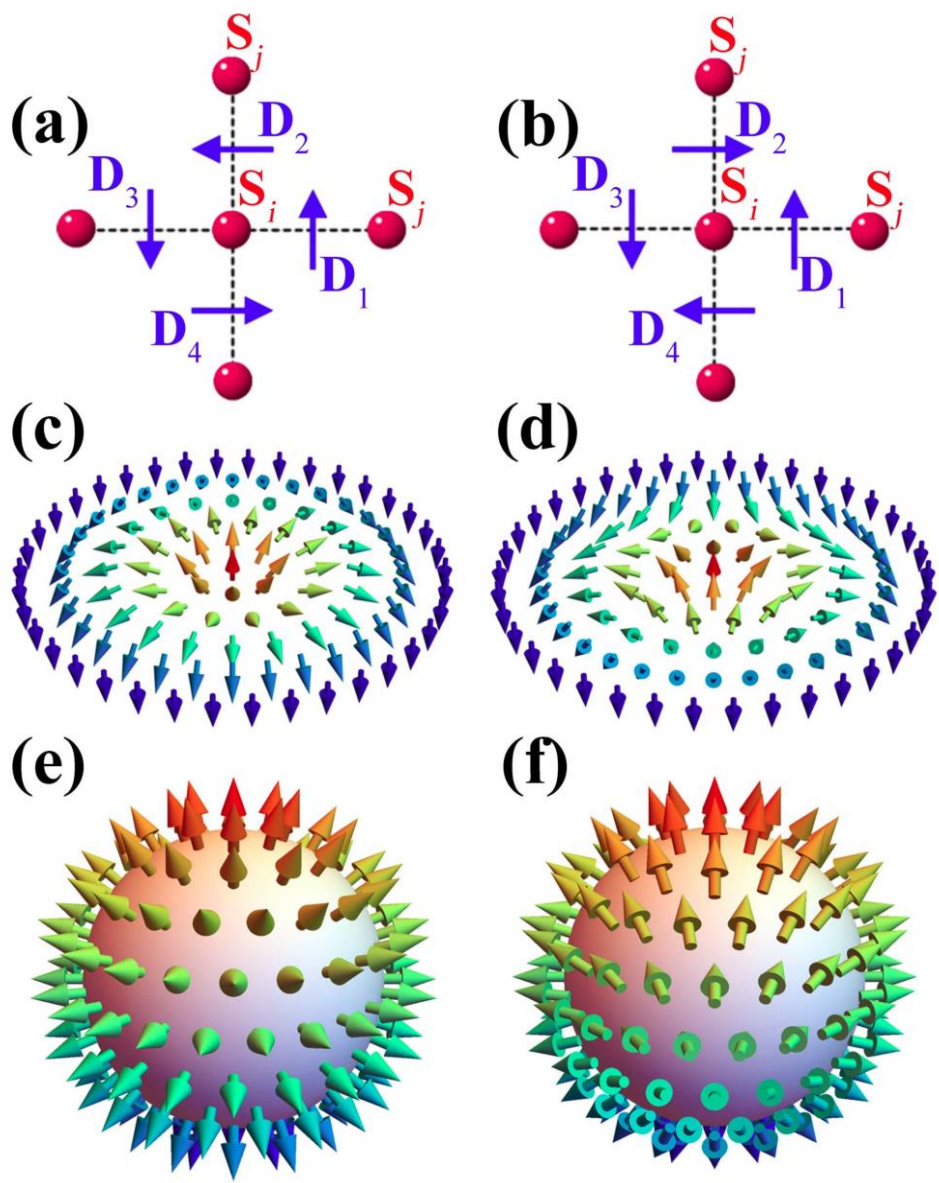

FIG. 1. Skyrmion and antiskyrmion configurations corresponding to isotropic and anisotropic DMI, respectively. (a) and (b), Schematic diagrams of isotropic and anisotropic DMI. Red balls indicate atomic spins at the interface while blue arrows indicate the DM vectors. (c) and (d), Arrows-array representation of skyrmion and antiskyrmion spin configurations. (e) and (f), Construction recipes for skyrmion and antiskyrmion in order parameter space. Note that panel c and d can be derived from panel e and f through stereographic projection. In panel c-f, different colors of the arrows correspond to different angles between the spins and normal direction of the interface.

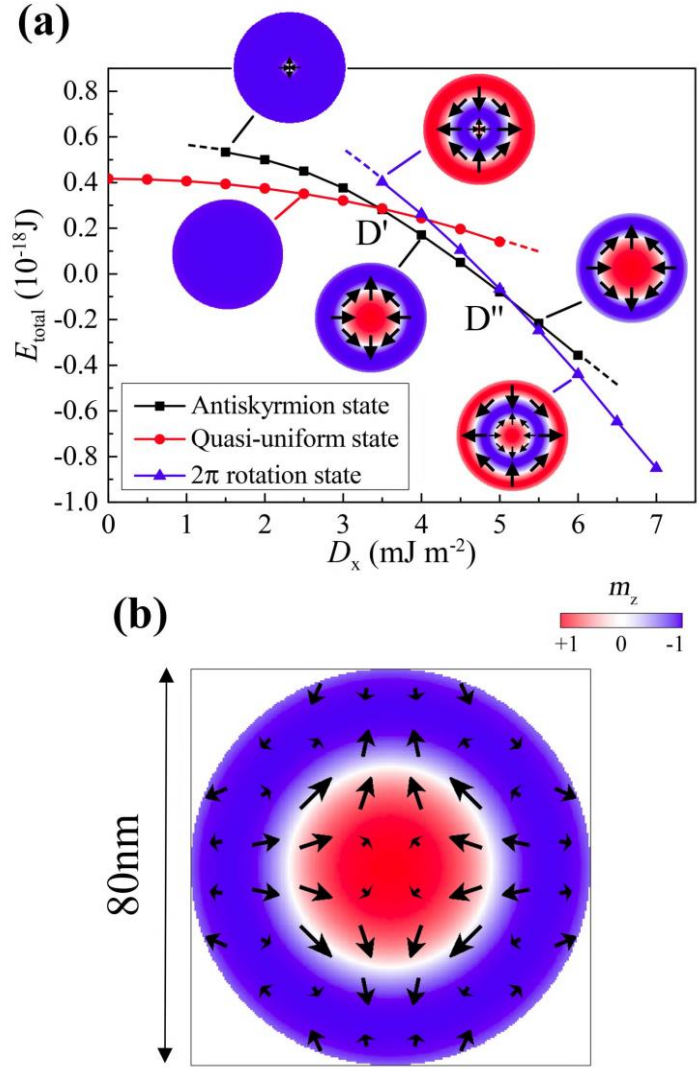

FIG. 2. Isolated antiskyrmion in a nanodisk (diameter, 80 nm) in presence of anisotropic DMI with $D_x = -D_y$. (a) Total micromagnetic energy (including the DMI, exchange, dipolar and anisotropy energies) for different states of the nanodisk as a function of $D_x$ along the line of $D_x = -D_y$ in $(D_x, D_y)$ space. Dashed lines indicate that the corresponding states are unstable in the simulations and tends to relax to more stable states. $D'$ and $D''$ indicate points where two different states are energetically degenerate. Insets show examples of relaxed magnetization distributions of the nanodisk for several points in the graph. (b) Magnetization distribution for an antiskyrmion ground state with $D_x = -D_y = 4$ mJ m$^{-2}$. Red, white and blue colours indicate out-of-plane magnetization distribution as shown in the color bar, as used throughout the paper. Black arrows highlight the in-plane orientation of spins.

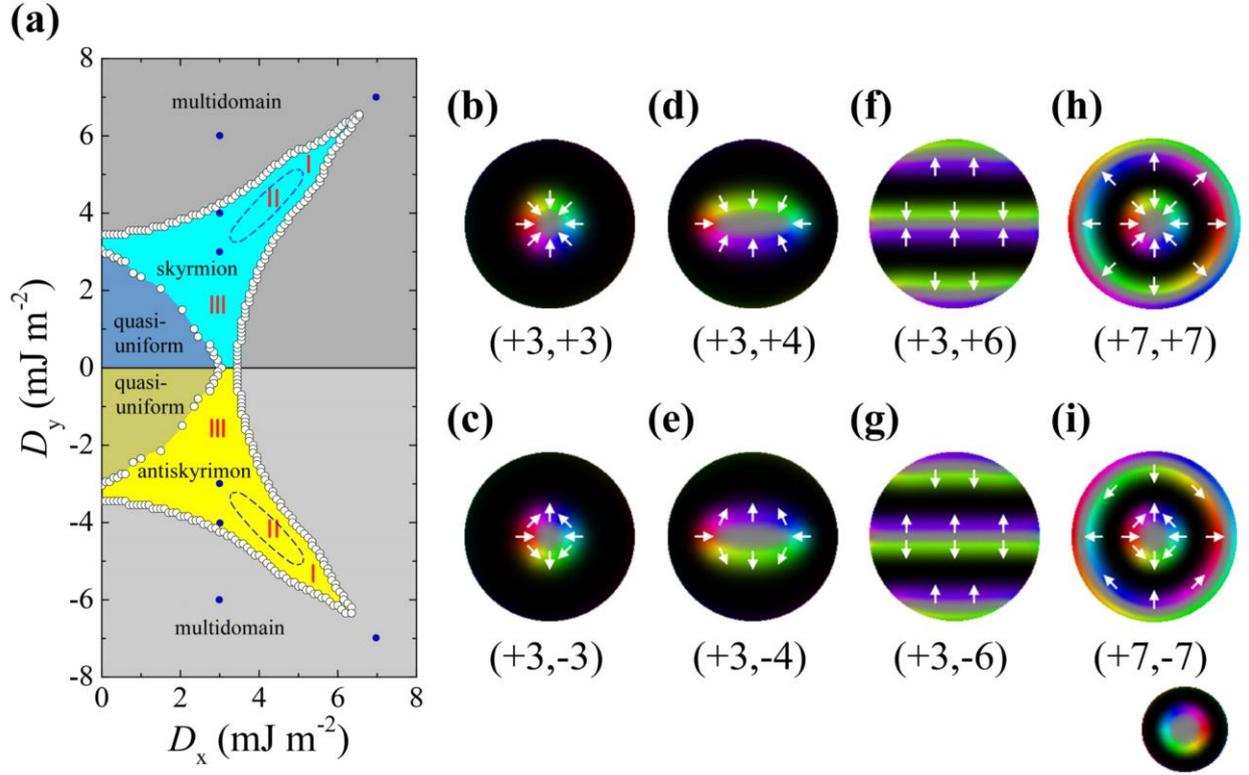

FIG. 3. Phase diagram of skyrmion and antiskyrmion states in $(D_x, D_y)$ space. (a) Phase diagram for a 80nm wide, 0.4nm thick cobalt nanodisk on a substrate inducing DMI in zero field. Regions coloured cyan, yellow and green correspond to quasi-uniform, skyrmion/antiskyrmion and multidomain states, respectively. White dots in a show the phase boundary obtained from the simulations. The blue dashed loops outline regions labeled II where skyrmion and antiskyrmions are the most stable states. In regions labeled I (III) the energy of multidomain (quasi-uniform) states is lower than that of skyrmion/antiskyrmion states. (b)-(i), Representative magnetization distribution graphs for $(D_x, D_y)$ values listed under each graph (in units of $mJ\,m^2$), blue dots in a mark the corresponding phase diagram coordinates. The relationship between different colors and in-plane magnetization directions is shown in the colour wheel, and in-plane spin orientations are highlighted by white arrow

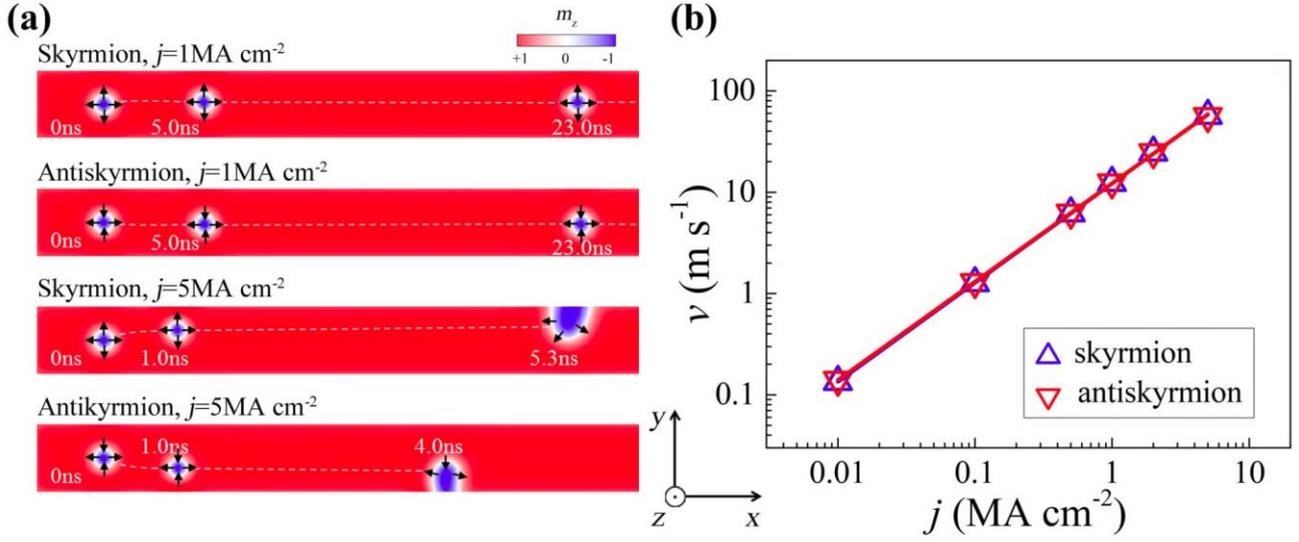

FIG. 4. Current-induced motion of skyrmions and antiskyrmions on nanotracks. (a) Current-driven motion of skyrmions and antiskyrmions on nanotracks for different current densities. White dashed lines indicates trajectories and black arrows show the in-plane spin orientations in the perimeters of the skyrmion/antiskyrmion. (b) The simulated Skyrmion and antiskyrmion velocity as a function of current density. The velocities of skyrmions and antiskyrmions are almost equal at all current densities.

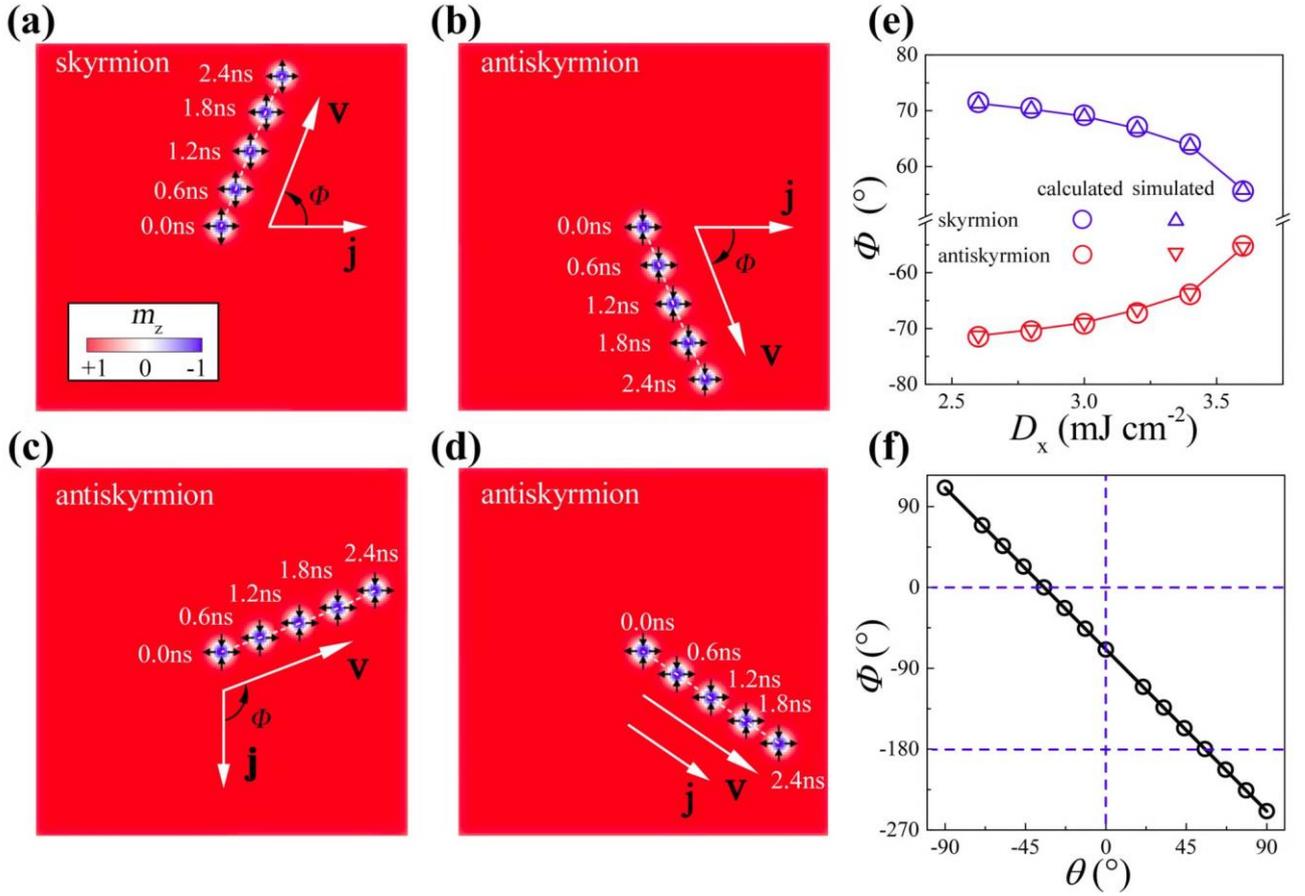

FIG. 5. Anisotropic antiskyrmion Hall effect. (a) Skyrmion motion trajectory with current along the x axis. (b)-(d) Antiskyrmion motion trajectories with the current direction at $\theta = 0°$, $-90°$ and $-34.5°$ to the x-axis, respectively. White dashed lines in panel a-d represent the trajectories. (e) Skyrmion and antiskyrmion Hall angles $\Phi$ as a function of DMI constant. (f) Antiskyrmion Hall angle $\Phi$ as a function of current direction $\theta$, solid line is a linear fit. In these simulations the size of the nanotrack is $300 \times 300 \times 0.4$ nm$^3$, and the applied current density is 10 MA cm$^{-2}$.

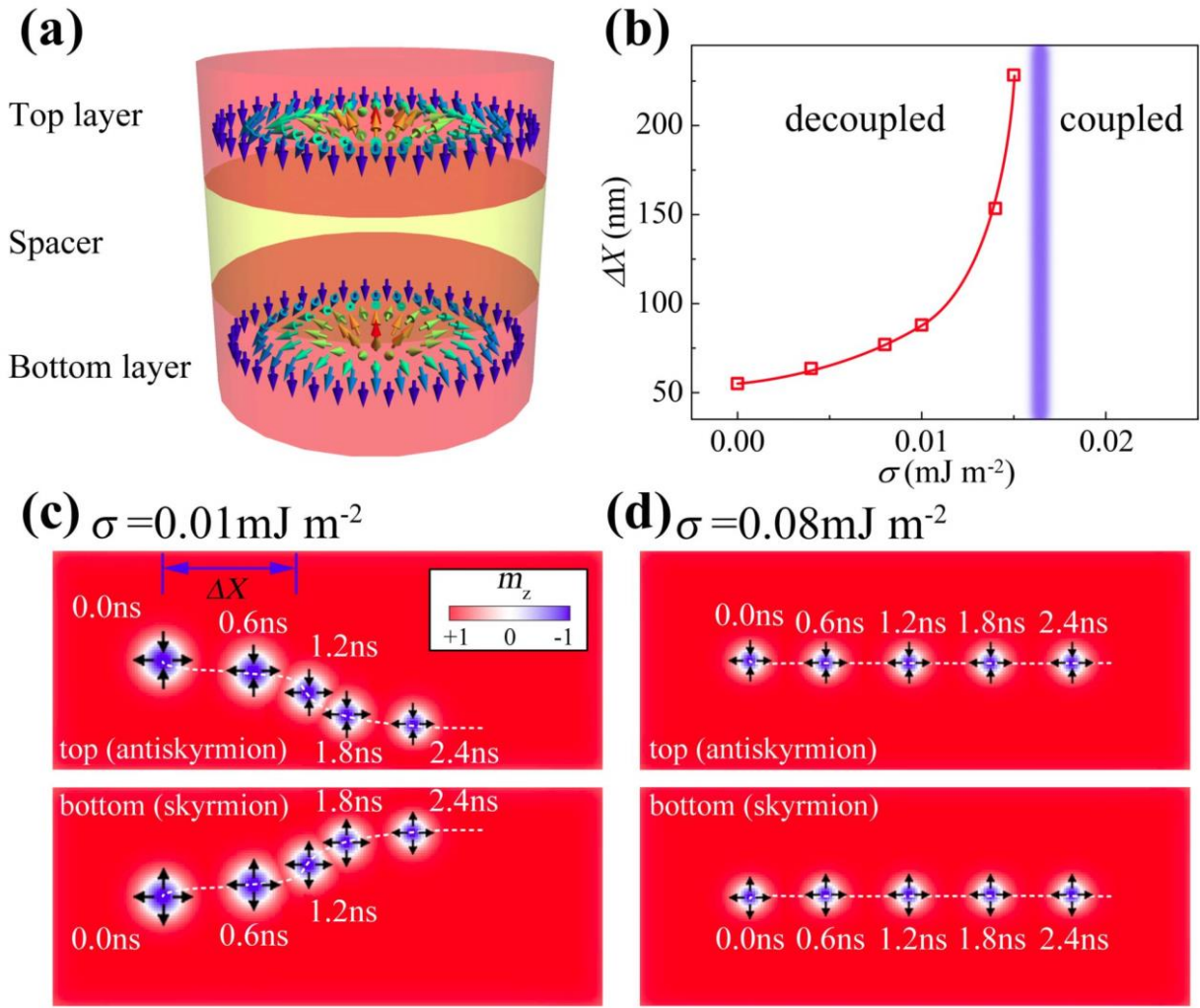

FIG. 6. Current-induced motion of ferromagnetically coupled antiskyrmion-skyrmion pairs. (a) Illustration of a antiskyrmion-skyrmion pair in a bilayer system. (b) decomposition distance ΔX as a function of the interface coupling strength $\sigma$. The red line is a guide to eye. The blue shadowed line represents the critical value of $\sigma$ above which the antiskyrmion-skyrmion pair remains coupled. (c) Current induced motion of antiskyrmion-skyrmion pair which is not sufficiently coupled, at coupling strength $\sigma = 0.01$ mJ $m^{-2}$. (d) Current induced motion of antiskyrmion-skyrmion pair that remains coupled, at sufficient coupling strength $\sigma = 0.08$ mJ $m^{-2}$. Dashed lines in panel c and d represent the trajectories. Size of the coupled bilayer nanotrack is 200×80×1.2 $nm^3$, current density is 5 $MA$ $cm^{-2}$, and $D_x = 3$ mJ $m^{-2}$.